\documentclass{moriond}

\def\Journal#1#2#3#4{{#1} {\bf #2}, #3 (#4)}

\def\PRL{\em Phys. Rev. Lett.}
\def\PRD{{\em Phys. Rev.} D}

\def\be{\begin{equation}}
\def\ee{\end{equation}}
\def\bea{\begin{eqnarray}}
\def\eea{\end{eqnarray}}

\newcommand{\Photo}{}

\begin{document}
\vspace*{4cm}
\title{Update on the BINGO 21cm intensity mapping experiment\footnote{We dedicate this update to the memory of our friend and colleague Prof Richard Davis OBE who passed away on 4th May 2016. He already played a role in the early stages of the  development of the BINGO project and we were looking forward to benefiting from his technical wisdom as part of the project.}}

\author{Richard Battye, Ian Browne, Tianyue Chen, Clive Dickinson,  Stuart Harper, Lucas Olivari, Michael Peel, Mathieu Remazeilles, Sambit Roychowdhury, Peter Wilkinson}

\address{Jodrell Bank Centre for Astrophysics, School of Physics and Astronomy, University of Manchester, Oxford Road, Manchester M13 9PL, U.K.}
\vskip -0.3cm

\author{Elcio Abdalla, Raul Abramo, Elisa Ferreira}

\address{Departamento de F\'{i}sica Matematica, Instituto de F\'{i}sica, Universidade de S\~{a}o Paulo, Rua do Mat\~{a}o 1371, S\~{a}o Paulo, Brazil}
\vskip -0.3cm

\author{Alex Wuensche, Thyrso Villela}

\address{Divis\~{a}o de Astrof\'{i}sica, INPE, S\~{a}o Jose dos Campos - SP, Brazil}
\vskip -0.3cm

\author{Manuel Caldas, Gonzalo Tancredi}

\address{Departamento Astronomia, Facultad de Ciencias, Universidad de la Republica, Igua 5225, 11400 Montevideo, Uruguay}
\vskip -0.3cm

\author{Alexander Refregier, Christian Monstein}

\address{Institute for Astronomy, Department of Physics, ETH Zurich, Wolfgang Pauli Strasse 27, 8093 Zurich, Switzerland}
\vskip -0.3cm

\author{Filipe Abdalla}

\address{Department of Physics and Astronomy, University College London, Gower St, London WC1E 6BT, U.K.}
\vskip -0.3cm

\author{Alkistis Pourtsidou}

\address{Institute of Cosmology and Gravitation, University of Portsmouth, Dennis Sciama Building, Burnaby Road, Portsmouth P01 3FX, U.K.}
\vskip -0.3cm

\author{Bruno Maffei}

\address{Institut d'Astrophysique Spatiale, Universite Paris Sud, 91405 Orsay Cedex, France}
\vskip -0.3cm

\author{Giampaolo Pisano}

\address{School of Physics and Astronomy, Cardiff University, Queens Building, The Parade, Cardiff CF24 3AA, U.K.}
\vskip -0.3cm

\author{Yin-Zhe Ma}

\address{School of Chemistry and Physics, University of KwaZulu-Natal, Westville Campus, Private Bag X54001, Durban 4000, South Africa}

\maketitle\abstracts{21cm intensity mapping is a novel approach aimed at measuring the power spectrum of density fluctuations and deducing cosmological information, notably from the Baryonic Acoustic Oscillations (BAO). We give an update on the progress of BAO from Integrated Neutral Gas Observations (BINGO) which is a single dish intensity mapping project. First we explain the basic ideas behind intensity mapping concept before updating the instrument design for BINGO. We also outline the survey we plan to make and its projected science output including estimates of cosmological parameters.}

\vfill\eject

\section{Introduction}

The estimation of cosmological parameters from the anisotropies of the Cosmic Microwave Background (CMB) within the standard $\Lambda$CDM model is now a mature subject~\cite{Planck}. However, very tight constraints are limited to the flat 6 parameter model; if the model is extended to include, for example, the dark energy equation of state parameter $w=P_{\rm de}/\rho_{\rm de}$ then the angular diameter degeneracy leads to only weak constraints on $w$ and increased error-bars on the other 6 parameters. In order to rectify this situation CMB data are often combined with a probe of Large-Scale Structure (LSS) and the most popular example is the use of the Baryonic Acoustic Oscillations~\cite{bao} (BAOs)  in the matter power spectrum deduced by performing galaxy redshift surveys in the optical waveband. The results of the BAO measurements are compatible with the CMB data from the {\it Planck} satellite and hence lead to tight constraints on any extension of the standard model. This is contrary to a number of other LSS probes~\cite{BCM} such as weak lensing, clusters counts and redshift space distortions.

In coming years the BAOs will be measured with increasing precision and over a wide redshift range by deep redshift surveys such as that which will be performed as part of the DESI project~\cite{desi}. The BAOs are expected to be a relatively clean probe of cosmology, but at the level of precision probed by future experiments this cannot be guaranteed. It is important to contemplate performing similar surveys in other wavebands and, for example, it has been suggested that spectroscopic surveys using the redshifted 21cm line of neutral hydrogen can be carried out by the SKA~\cite{abdalla}. This will require the significant collecting area of the SKA to detect high redshift galaxies due to the long lifetime of the spin-flip transition responsible for the line.

\section{21cm intensity mapping}

Observations made by a radio telescope are diffraction limited in the plane perpendicular to the line-of-sight and hence a telescope with a diameter of 10s of meters will have resolution $\sim 1\,{\rm deg}$, but a spectral resolution of $\sim 1\,{\rm MHz}$ around an observation frequency of $\sim 1\,{\rm GHz}$ is easily achievable. Such a telescope will be unable to detect the 21cm signal from anything but the most closeby galaxies. However, there will be an overall integrated signal due to the combined effect of all galaxies within the beam, and this will vary from position to position on the sky. If this signal can be extracted from the much brighter continuum emission then processing it will allow an estimate of the matter power spectrum and techniques similar to those used in the optical can be used to extract cosmological information such as the BAO scale. This is the basic idea of 21cm intensity mapping~\cite{BDW,Peterson}.

In order to estimate the size of the signal consider a cell of the Universe corresponding to $(30\,{\rm arcmin})^2\times 1\,{\rm MHz}$ observed at a frequency of $1\,{\rm GHz}$, which corresponds to $z\sim 0.4$. The volume of the cell is $(8 h^{-1}\,{\rm Mpc})^3$, such a region will contain an average of $\sim 10^{10}\,M_{\odot}$ neutral hydrogen and the r.m.s. variance in the density on such scales is $\sigma_8\approx 1$. The average brightness temperature is $\sim 100\,\mu{\rm K}$ and such a signal could be detected with a radio telescope with instantaneous sensitivity on a bandwidth of $1\,{\rm MHz}$ of $\sim 50\,{\rm mK}\,{\rm s}^{1/2}$  in 2-3 days of integration suggesting that the detection of such a signal over a significant portion of the sky would be possible in around a year observing.

\begin{figure}
\centerline{\includegraphics[width=1.0\linewidth]{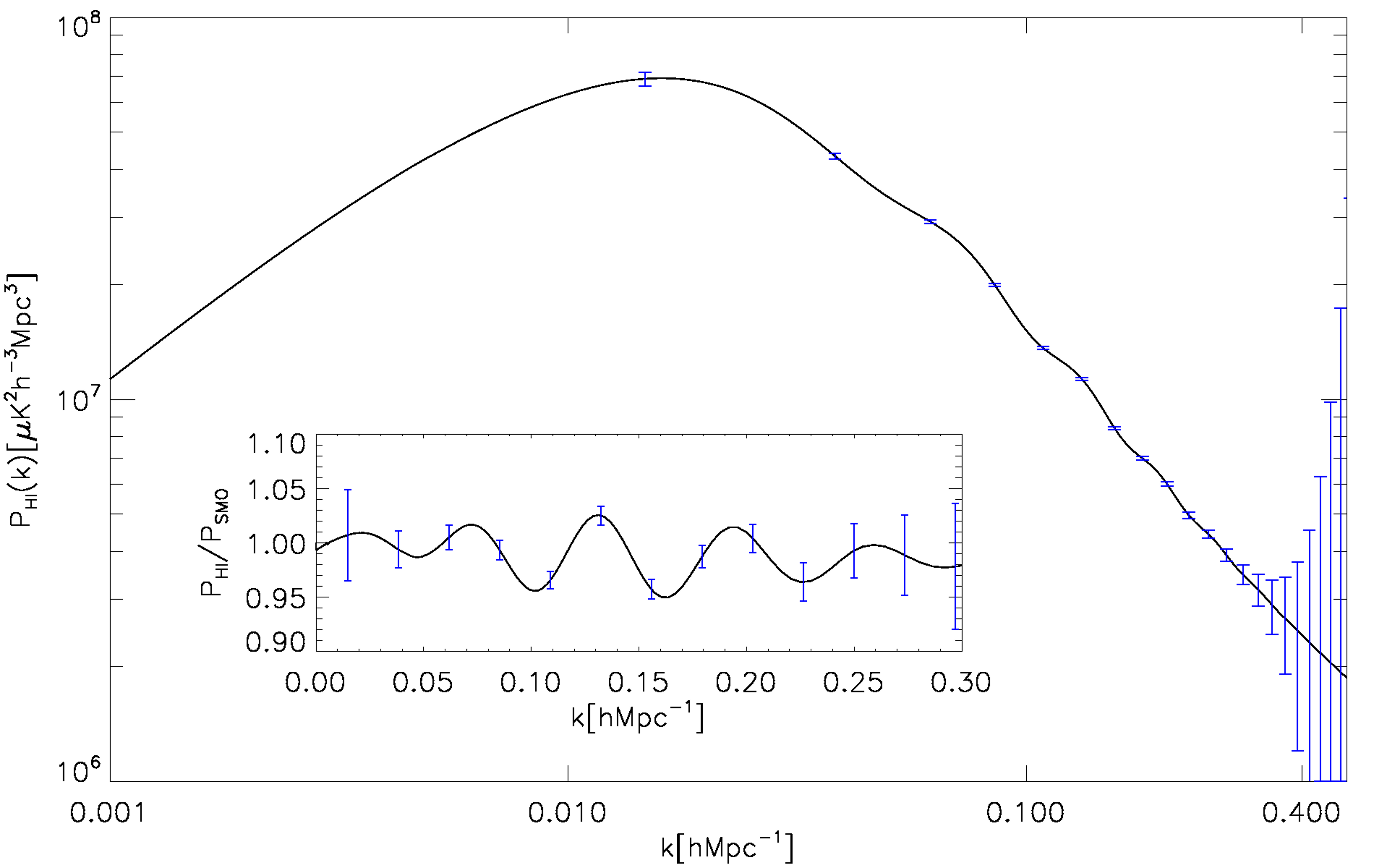}}
\caption[]{The HI power spectrum, $P_{\rm HI}(k)$, versus wavenumber, $k$, at $z=0.28$ - corresponding to  the centre frequency of BINGO - along with projected errors on the reconstruction from the proposed BINGO survey with 50 feedhorns and $1\,{\rm year}$ of on-source integration time covering $2000\,{\rm deg}^2$. Included as an inset is the power spectrum divided by a``smooth spectrum" to isolate the BAOs.}
\label{fig:ps}
\end{figure}

In fig.\ref{fig:ps} we plot the expected 3D power spectrum as a function of wavenumber, $k$, for a redshift bin defined by frequencies $960-1260\,{\rm MHz}$. We also present the BAO signal, highlighted by the ratio of the actual 3D  power spectrum and one which is smooth, against wavenumber $k$. The amplitude of the signal is compatible with our earlier estimate and we see that the BAO signal is localised in the range $k=0.02\,h^{-1}\,{\rm Mpc}$ to $0.2\,h^{-1}\,{\rm Mpc}$.

A number of dedicated projects have been proposed to detect this signal. These include interferometers such as CHIME~\cite{chime} and TIANLAI~\cite{tianlai}, and also single dishes with multiple feedhorns such as BINGO (which stands for BAO from Integrated Neutral Gas Observations). The two approaches are complementary since interferometers are typically targeted at high redshifts ($z\sim 0.8$), while single dishes are more suited to lower redshift ($z\sim 0.3$) - the reason being that it can become prohibitively expensive to achieve the resolution necessary to resolve the BAO scale at higher redshifts (lower frequencies) using a single dish. Ultimately it might be possible to perform an intensity mapping survey using the SKA using both interferometric and single dish modes and such a survey could be competitive with the {\it Euclid} satellite~\cite{bull}.

\section{Updated BINGO concept}

The basic idea of BINGO was discussed in the conference proceedings of the 2012 Moriond Cosmology meeting~\cite{bingo1} and the single dish idea for intensity mapping has been studied in detail~\cite{bingo2}. In this contribution we present an update of the design and the projected science output.

\subsection{New optical design and basic telescope/survey parameters}

The underlying concept behind the project is to provide a simple, clean and cheap approach to 21cm intensity mapping. We will build on the experience of measuring the anisotropies in the Cosmic Microwave Background (CMB) which has many similarities, although measurements of the CMB are at much higher frequencies and foregrounds are likely to be much more of a problem in the case of intensity mapping. 

\begin{figure}
\centerline{\includegraphics[width=0.7\linewidth]{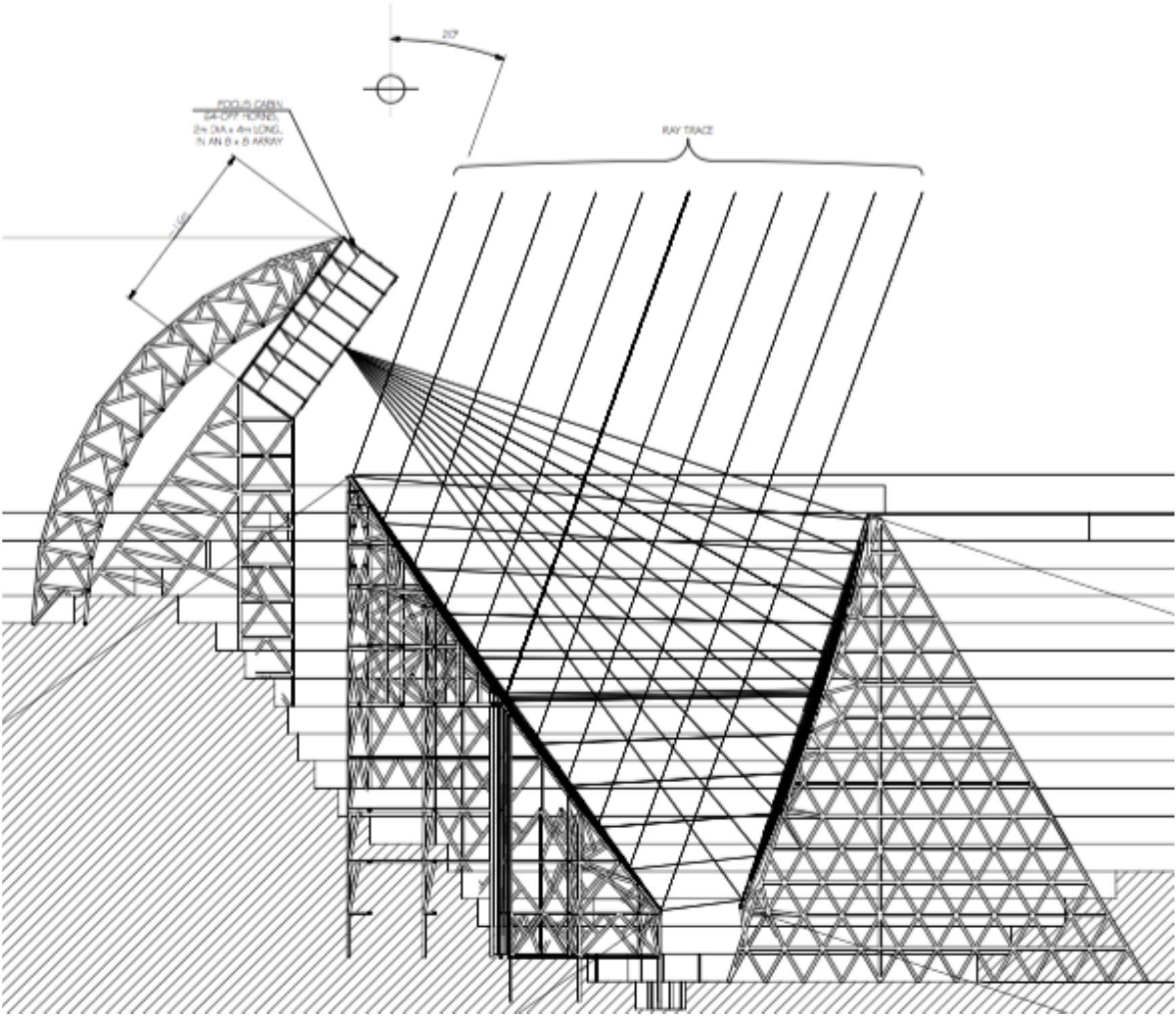}}
\caption[]{Basic optical design of the BINGO telescope from side-on with a draft sketch of the mechanical structure. On the left is the focal plane array with rays focussed on a single pixel and on the right is the secondary mirror that has diameter $\sim 40\,{\rm m}$ which focusses the rays from the the primary mirror of diameter $\sim 48\,{\rm m}$.The primary is in the middle and this reflects the rays from the sky onto the secondary mirror.}
\label{fig:optical}
\end{figure}

Originally we suggested~\cite{bingo1,bingo2} using a ``cliff" telescope with a static dish at the bottom of a cliff of height $\sim 100\,{\rm m}$ and a focal ratio $\sim 3$ in order to accommodate  a large focal plane array with very little aberration at the edges. This would have been located at the top of the cliff which would allow the very heavy horns required to be supported by the cliff rather than an expensive mechanical structure. Unfortunately we were unable to find an appropriately sized cliff at the optimum latitude $\sim \pm (30-40)^{\circ}$ and therefore we have decided to evolve our the design to have two dishes of comparable size with one acting as a secondary - the so-called Crossed-Dragone/Compact Range Antenna design often used in CMB polarization experiments. The basic optical design, which is more compact than the original one, is illustrated in fig.~\ref{fig:optical}.

The BINGO telescope will comprise of two static dishes surfaced with wire mesh, each with a diameter of around 40~m. They will be significantly under-illuminated ($\sim$25~m) in order to suppress sibe-lobes and the beam will have a full-width-half-maximum (FWHM) of $\theta_{\rm FWHM}\approx 40\,{\rm arcmin}$ at an observing frequency of $1\,{\rm GHz}$ corresponding to a  wavelength of $\lambda\approx 30\,{\rm cm}$. The telescope structure and the focal plane array of $\approx 50-100$ horns - will be supported by the sides of a quarry (see discussion of the site below). The receiver system will be a pseudo-correlation design in order to suppress the $1/f$ noise and achieve a ``knee'' frequency of $\sim 1\,{\rm mHz}$ over a  bandwidth of $1\,{\rm MHz}$. The output of each receiver chain  will be the difference between the signal received from a main horn pointing at the dish and a reference horn which will point directly at the South Celestial Pole (SCP).  There will need to be as many reference horns as main horns, and each will have a beam size $\sim 25\,{\rm deg}$. The overall system temperature, $T_{\rm sys}$, of this uncooled system, built from ``off-the-shelf'' RF components, will be $\approx  50\,{\rm K}$ and have an overall instantaneous bandwidth of $\Delta f_{\rm inst}=300\,{\rm MHz}$ from $f_{\rm obs}=960\,{\rm MHz}$ to $1260\,{\rm MHz}$ corresponding to a redshift range of $z\approx 0.13-0.48$. The basic properties of the telescope are presented in table~\ref{tab:parameters}.

The telescope will be located at latitude $\approx -32^{\circ}$ and perform a drift scan survey at declination $\approx -15^{\circ}$. The receivers will be arranged so as to create an instantaneous field of view with a width of $15\,{\rm deg}$ in the direction perpendicular to the scan and up to  $15\,{\rm deg}$ in the direction of the scan dependent on the number of horns. This will facilitate a survey of $15\,{\rm deg}\times 360\cos(15^{\circ})\,{\rm deg}$. The reference horns will point at the SCP, which will provide a constant signal with a similar spectrum to the rest of the sky enabling a good balancing of the pseudo-correlation system that will lead to cancellation of the $1/f$ noise. Any residual drifts in the receiver baseline will be removed by a combination of calibration and component separation techniques, for example, using Principal Component Analysis~\cite{MA} (PCA) or Generalised Needlet Internal Linear Combination~\cite{lucas} (GNILC).

\begin{table}
 \caption{Summary of proposed BINGO telescope parameters}
 \vspace{0.4cm}
 \begin{center}
 \begin{tabular}{|c|c|}
 \hline
Main reflector diameter & 40~m \\
Illuminated diameter & 25~m    \\
Angular resolution FWHM for observing at $1\,{\rm GHz}$) & $\sim$ 40 arcmin   \\
Number of feeds &50-100     \\
System temperature & $\approx 50\,{\rm K}$\\
Instantaneous field of view (dependent on number of horns) & 15\,${\rm deg}\times 15\,{\rm deg}$ \\
Frequency range & $960\,{\rm MHz}$ to $1260\,{\rm MHz}$ \\
Number of frequency channels & $\geq$300  \\
 \hline
 \end{tabular}
 \end{center}
 \label{tab:parameters}
 \end{table}

\subsection{Foam horns}

The BINGO design is extremely simple but there is one significant practical hurdle that needs to be overcome. The horns will be extremely large - they will need to have a diameter of $\sim 1.7\,{\rm m}$ and a length of $\sim 4.5\,{\rm m}$ and hence each horn would weigh around $\sim 2\,{\rm tonnes}$ if made from a conventional metal design. The focal plane array of 50-100 horns would be prohibitively expensive and  require an extremely solid - and hence expensive structure - to mount them. 

\begin{figure}
\centerline{\includegraphics[width=0.5\linewidth]{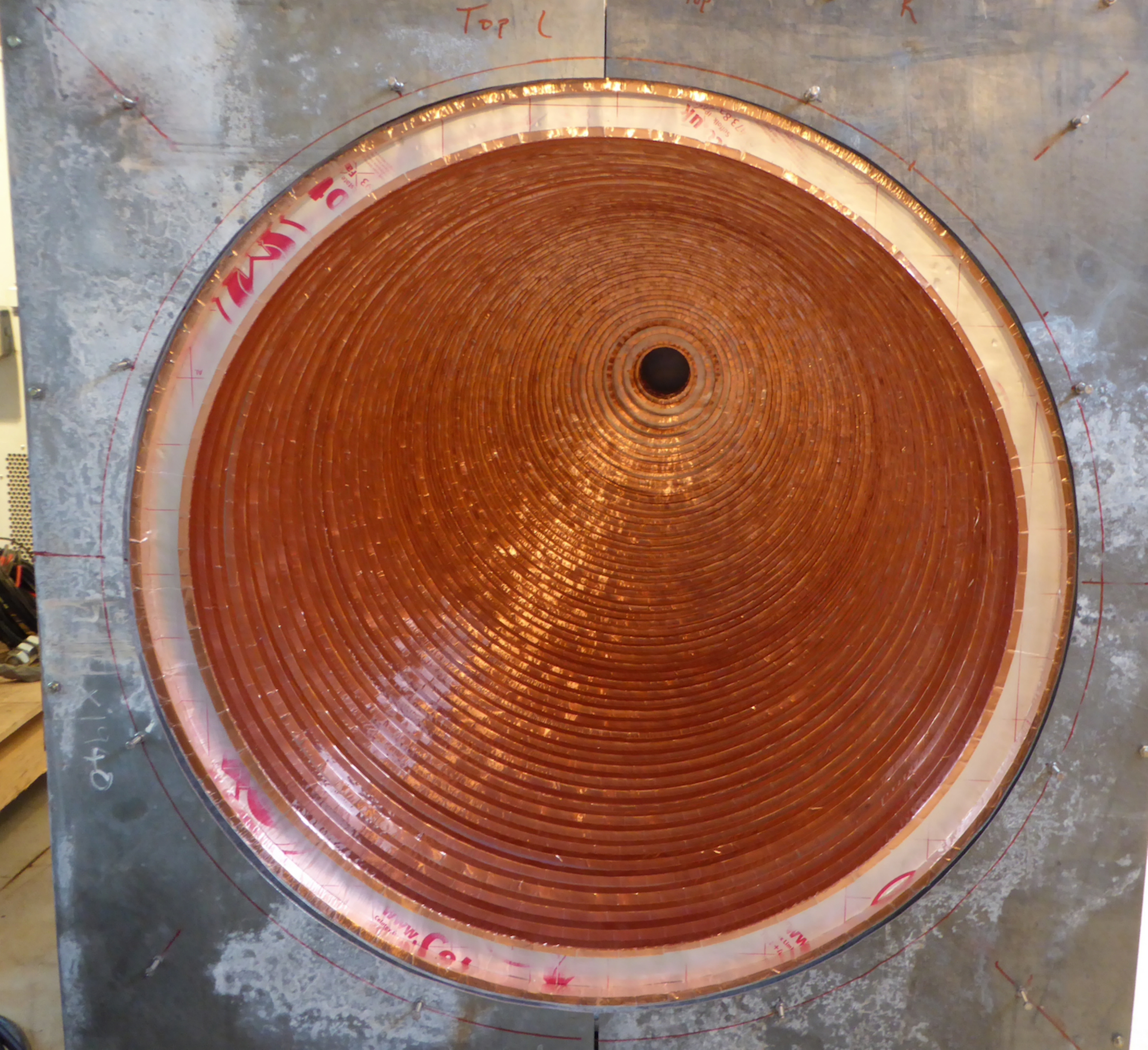}\includegraphics[width=0.5\linewidth]{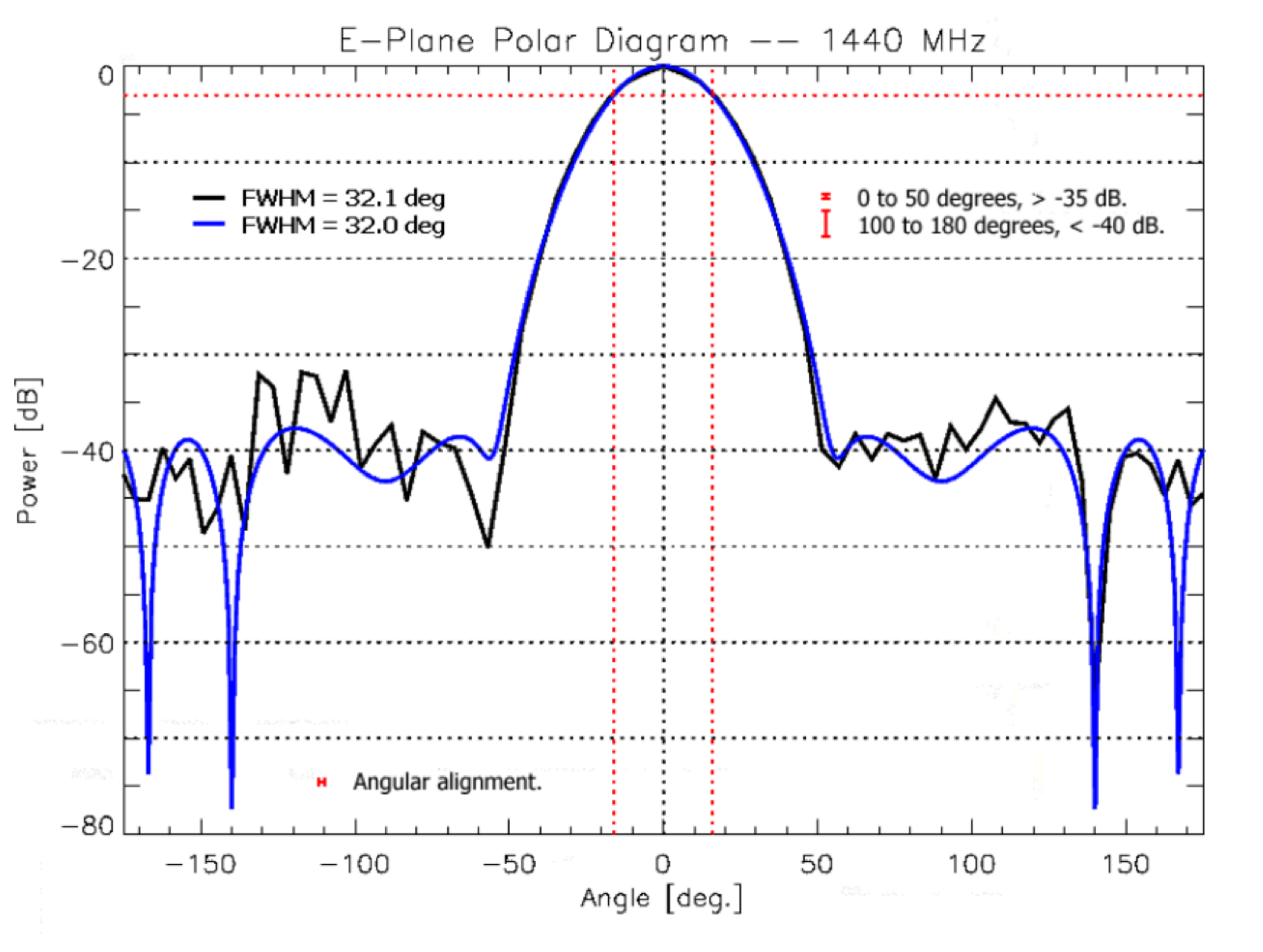}}
\caption[]{On the left is a picture looking down the throat of the large foam based horn described in the text and on the right is the polar diagram of the horn~\cite{horn}. Notice the good agreement, at least for the main beam between the measured beam and the theoretical prediction. Each indentation is a foam sheet covered with copper tape.}
\label{fig:horn}
\end{figure}

In order to overcome this problem we have designed corrugated horns made from foam that are coated with copper tape, which will be significantly lighter and less expensive. These horns are made  of sheets of foam that are $\sim 2.5\,{\rm cm}$ thick which have an appropriate sized circular hole cut in them and copper tape stuck on. The sheets are then compressed together in order for the horn to have good electrical conduction. In fig.~\ref{fig:horn} we show a 124-element horn and the measured polar-diagram for the beam profile compared to a theoretical profile. The agreement between the measurement and theory is excellent in the high-signal-to-noise region for the measurements around the main beam and the side-lobes are also satisfactorily reproduced within the measurement accuracy. Note also that the sidelobes are reduced by  -40dB relative to the main beam.

\subsection{Site in Northern Uruguay}

We have selected a site in Northern Uruguay close to the small town of Minas   Corales. It will be in the Castrillon quarry - a dis-used gold mine that is well-matched to the BINGO optics being $\approx 50$\,m deep and $\approx 40$\,m wide, with a $\approx 45^{\circ}$ slope at one end, orientated close to N-S (see fig.~\ref{fig:site}).  The main advantage of the site is the low population density of the area resulting in low levels of Radio Frequency Interference (RFI). Initial RFI tests in 2013 have shown that the $960-1260$\,MHz band is relatively clean at the site. We also have support from the mining company and the local community.

\subsection{Project status and timeline}

A proposal has been submitted to FAPESP,  the S\~{a}o Paulo regional funding agency, to cover a significant fraction of the capital costs and it is in the final stages of being approved subject to (i) the successful construction of a full proto-type receiver chain including a full-scale horn,  (ii) a design study of the telescope structure, and (iii) support is forthcoming from the other relevant funding agencies.

Phase 1 of the project which involves fulfilling the criteria (i)-(iii) is underway and will be completed by the middle of 2017.  It is hoped that phase 2 of the project - the construction of the telescope within the Castrillon quarry - will take place during 2017-2018. The aim is to start science operations toward the end of 2018 and to take data for at least 4 years.

\section{Projected science output and component separation}

The main goal of BINGO is a measurement of the BAO scale in the redshift range $z=0.13-0.48$ and the design has been optimised for this~\cite{bingo2}. The choice of $40\,{\rm arcmin}$ resolution was chosen in order to achieve the best constraint on the BAO scale, $k_{\rm A}$ without getting into the situation where any improvement in the constraint leads to a significant the extra cost due to the increased resolution. Using a thermal noise dominated survey with 50 horns and 1 year of on-source integration time covering $\approx 2000\,{\rm deg}^2$, it was shown that a constraint of $\delta k_{\rm A}/k_{\rm A}\approx 0.025$ could be achieved. With an increased number of horns, that might be possible subject to cost, it should be possible to do somewhat better than this.

We have simulated the reconstruction of the power spectrum and the expected cosmological parameter estimation from the proposed BINGO survey in conjunction with information coming from the CMB observations made by {\it Planck}. Projected errorbars on the power spectrum for the survey described above are included in fig.~\ref{fig:ps}. CMB observations by themselves cannot constrain the equation of state of dark energy, $w$, but we find that a combination of {\it Planck} and BINGO can constrain $\Delta w\approx 0.1$.

\begin{figure}
\centerline{\includegraphics[width=1.0\linewidth]{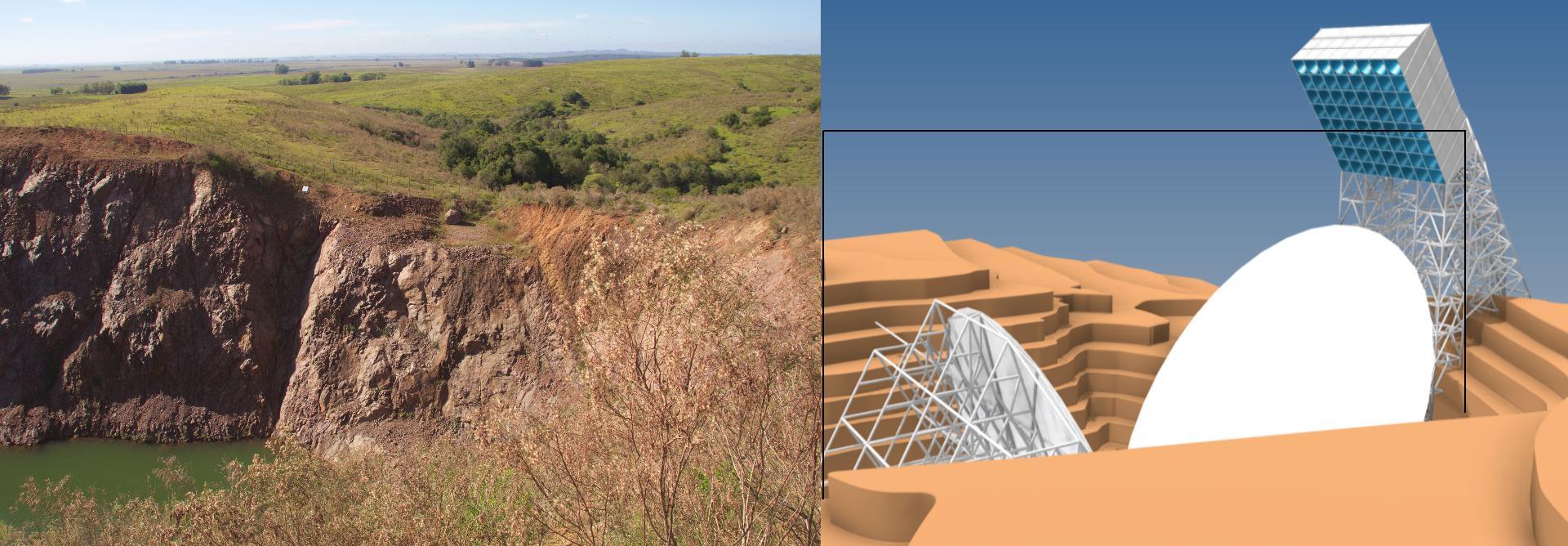}}
\caption[]{On the left is a picture of the Castrillon quarry which is the proposed site of the BINGO telescope. On the right  is an engineering drawing of the telescope with primary, secondary and focal plane array of horns in position. The black box in the right hand figure is approximately the plane of the picture on the left.}
\label{fig:site}
\end{figure}

In addition to the measurement of BAOs there are many other interesting scientific goals that will be pursued as part of the BINGO project.

\begin{itemize}

\item {\it Neutrino masses:} The shape of the matter spectrum is sensitive to the masses of the neutrinos. If there is no significant scale dependent bias between the  HI power spectrum and the underlying matter power spectrum then it will be possible to measure the sum of the masses of the neutrinos, $\sum m_{\nu}$, using BINGO observations.

\item {\it Redshift space distortions:} Redshift space distortions can can be extracted from the 21cm intensity mapping signal by splitting the power spectrum into components which is perpendicular and parallel to the line-of-sight. It has been shown that such observations have the capacity to distinguish between different models of modified gravity~\cite{HBC}. Observations made by the BINGO telescope will be able to probe such models.

\item {\it Cross-correlation with optical redshift surveys:} It should be possible to cross-correlate the 21cm intensity mapping survey provided by BINGO with the redshift survey provided by the Dark Energy Survey. We have estimated that there will be $\sim 800\,{\rm deg}^2$ of overlap between the two surveys. This should lead to additional constraints on the origin of cosmic acceleration~\cite{alkistis}.

\item {\it Fast Radio Bursts (FRBs):} The Fast Radio Burst (FRB) phenomenon is arguably the most exciting and unexpected astronomical discovery so far this century~\cite{FRB}. BINGO with its multiple beams giving wide instantaneous sky coverage will be a great engine for FRB discovery. At present less than 20 examples are known and BINGO with suitable digital processing should be able to find them at a rate of tens per year. In order to achieve this, the BINGO telescope will need to be fitted with additional ``backend" electronics needed to achieve a higher time resolution than necessary for intensity mapping.  In an ambitious third phase of BINGO, small outrigger telescopes could be added in order to measure real-time accurate positions for FRBs to enable optical identification of their hosts, a step essential to distinguish between the many proposed explanations of the origin of FRBs. 

\item{\it Galactic science:} By virtue of the drift scan strategy the BINGO survey will cross the 
Galactic plane. Regions where the Galactic emission is particularly high will be excluded from the cosmological analyses, but they will provide interesting information on the nature of the this Galactic emission which will be combined with the observations, for example, from the {\it Planck} satellite to provide an improved model of the galaxy. Interesting information that could be gleaned from this will include line emission from the radio-recombination lines (RRLs) due to the ionised gas in the Galactic plane~\cite{alves}.

\end{itemize}

\section*{Acknowledgments}
We would like to thank Adrian Galtress for his work on the mechanical design of the BINGO telescope and the engineering drawings presented in fig.~\ref{fig:site}.

\end{document}